# Startup Growth Hacking – a Board Game Approach

Kai-Kristian Kemell[1], Polina Feshchenko[1], Joonas Himmanen[1], Abrar Hossain[1], Johannes Impiö[1], Furqan Jameel[1], Joni Kultanen[1], Raffaele Luigi Puca[1], Juhani Risku[1], Anssi Sorvisto[1], Teemu Vitikainen[1], and Pekka Abrahamsson[1]

[1]University of Jyväskylä, 40014 Finland
{kai-kristian.o.kemell | pekka.abrahamsson}@jyu.fi

**Abstract**

*Startups seek to create highly scalable business models. For startups, growth is thus vital. Growth hacking is a marketing strategy advocated by various startup practitioner experts. It focuses on using low cost practices while utilizing existing platforms in creative ways to gain more users for the service. Though topics related to growth hacking such as marketing on a general level have been extensively studied in the past, growth hacking as a practitioner-born topic has not seen much interesting among the academia. To both spark interest in growth hacking, and to facilitate teaching growth hacking in the academia, we present two board games intended to serve as an engaging introduction to growth hacking for students.*

## 1 INTRODUCTION

Though most companies are concerned with growth, for startups growth is often far more vital than it is for more mature organizations. In fact, some definitions incorporate a highly scalable business model as one characteristic of what a startup is and how it differs from traditional, smaller business organizations [17].

Strategies for growth are various. In terms of growing through user or customer acquisition, *marketing* is a key activity. Marketing strategies are numerous, ranging from e.g. digital viral marketing to traditional forms of display advertising done through television advertisements. As startups operate under a notable lack of resources, especially financially, and especially early on in their life cycles, this often limits their options in choosing marketing strategies. *Growth hacking* is a marketing strategy that focuses on low cost practices and using existing platforms in creative way, and is thus well-suited for startups in this sense.

Currently, growth hacking has seen little interest in the academia. Though marketing is a long-standing area of research in economic disciplines, and search engine optimization (SEO) and other areas of research closely related to growth hacking have been extensively studied in the field of information technology, growth hacking has not been directly studied. As a result, startup education in the academia currently has little scientific basis to build on. Thus, while the importance of growth is well known, methods for teaching growth hacking in the academia are lacking.

To facilitate teaching growth hacking, we present two educational growth hacking board games in this paper. These games were developed as a part of a course on growth hacking and based on a multi-vocal literature review on the topic. Together, they serve as an introduction to the topic while also teaching various actionable growth hacking practices.

The rest of the paper is organized as follows. The next section discusses growth hacking as a construct and in practice through both practitioner and academic literature on the topic. In the third section we discuss the course design of the course during which the games were developed. Then, in the fourth section, we present the two board games, before concluding the paper in the fifth and final section where we also discuss the games and the course.

## 2 GROWTH HACKING

The construct *growth hacking* was popularized by practitioner expert Sean Ellis [12] in his blog about startup marketing[1]. It has remained a construct primarily used by practitioners and has seen little traction in the academia. However, various topics closely related to growth hacking, such as marketing in general and different SEO techniques, have been studied in both IT and economic disciplines.

---

[1] http://www.startup-marketing.com/

Growth hacking is a marketing strategy [10]. As the name implies, it is about using various growth hacking techniques or practices to "hack" the growth of a company, often a startup. In practice, growth hacking is technology-oriented and relies on using technical practices, with one of the main tasks of a so-called growth hacker in fact being (software) development [10]. This, Herttua et al. [10] underline, is one of the main differences between growth hacking and other marketing strategies such as viral marketing or guerilla marketing.

In academic literature thus far, the following characteristics have been associated with growth hacking [10, 12]:
- Use of data in the form of metrics
- Changing the service based on data
- Low cost practices
- "Pulling" users to the service as opposed to "pushing" the service to them
- Using existing platforms in creative ways
- A/B testing

As a technical form of marketing, growth hacking combines marketing, statistics, and development. A growth hacker would need to have the skills required to make changes to the software service based on the data he collects. While this could be accomplished by a team of individuals, each versed in one of the required skills, the startup focus of this construct typically focuses on growth hacking carried out by a single "growth hacker" in an organization. This is related to the lack of resources associated with startups: hiring a team to do the job of one expert is not feasible for a startup. [11]

In practice, growth hacking is carried out via various growth hacking practices, which are often referred to as "techniques", "tactics", or simply growth "hacks" by practitioners [5]. Practices that can be considered growth hacking depending on the definition given to it are seemingly numerous. They range from social media practices such as following individuals or organizations in hopes of gaining followers in return, to sales-related practices such as offering free software trials or downselling upon subscription cancellation.

A famous and commonly cited example of growth hacking in practice is the story of Hotmail [5, 11]. To tackle their growth issues early on, Hotmail implemented the signature text "PS. I Love You. Get your free e-mail at Hotmail" into all e-mails sent from their service. Having tried various other forms of marketing, this proved to be far more effective. Following the I Love You campaign, Hotmail quickly grew from a few thousand users to a few million users and sold its service to Microsoft not long afterwards. In this fashion, growth hackers aim to "hack" the growth of a service or company by both being creative with existing platforms and using low cost practices to drive growth.

## 3 RESEARCH DESIGN

The board games presented in this paper were created during the course "TJTS5792 Advanced Lean Startups" in the University of Jyväskylä. The games were developed by two teams of Information Systems (IS) students, under the supervision of the teaching staff of the course. For creating the board games, we conducted a multi-vocal literature review on growth hacking prior to the start of the course. As we discovered that the construct had not seen much traction in the academia at the time of writing, the review ultimately focused on practitioner literature.

Because this is an educational paper, we will also go over the course design on a more general level in this section. The first sub-section covers the pre-course activities of the teaching staff (the multi-vocal literature review) and the creation of the board games. The second sub-section discusses how the rest of the course was structured.

### 3.1 The Literature Review

In order to discover relevant practitioner literature, we conducted a set of Google searches, and utilized book review site reviews. The search string used was "growth hacking book".

As these board games were produced as a part of a university course, we limited ourselves to book material only as far as practitioner literature was concerned. This choice was made to provide the students with clear reading materials, one book per student. Furthermore, gray literature such as blogs and non-peer-reviewed online articles were not included into this review as we wanted to ensure, to some extent, the quality of the reading materials. Though a published book is no indication of correct facts, books do go through quality control practices, highly varied as they may be.

First, the context of the word match was examined: how was the construct "growth hacking" used in the book? Was it simply mentioned in passing, largely unrelated to the topic at hand, or were growth hacking themes actually discussed? Secondly, the contents of the book were screened based on the table of contents. Did the book discuss growth hacking themes? I.e. acquiring users or customers through means that could be classified as growth hacking.

After the content screening, a quality appraisal process was conducted using public book reviews. For each resulting book, either Good Reads or Amazon Reviews were used to conduct a second quality appraisal. Only books rated 3.8 or higher on a scale of 0 to 5 were included into the list of books to be reviewed. Once the books had been selected, they were included in the course readings.

### 3.2 Course Design

The course was five weeks long. Each week, there was a lecture and the students were given an assignment to be completed for the next one. The course structure is outlined in the table below (Table 1).

| Week | Tasks |
|---|---|
| 1 | Lecture. Each student assigned one book to read for the next session and to summarize it through a presentation. |
| 2 | Lecture. Students split into two teams, each tasked with creating a board game on growth hacking. |
| 3 | Brief lecture. Board games played. Course participants, including mentors, split into pairs to create educational videos on growth hacking. |
| 4 | Lecture. The same pairs of participants created more videos and prepared a plan for utilizing the practices. |
| 5 | Lecture. Practices utilized in a real setting. The participants presented their growth hacking practice(s) and their lessons learned in the course end event. |

In order to create the board games, the students were first tasked with reading one book each after the first lecture. After reading the book, each student was to teach its contents to the other students through a presentation[2]. Following the presentations during the second week of the course, the students split into two teams. Both teams were to create a board game on growth hacking as a team, in one week. They were not given explicit guidelines on how to categorize the growth hacking practices, or what to focus on. They were simply tasked with creating a game they felt taught growth hacking to its players, focusing on whatever they felt was important based on the readings.

After the student teams had created the board games, we played the games during the sub-sequent course meeting on the third week. Then, based on the learnings from both the books and the games, we wanted the students to utilize the growth hacking skills and practices in a real setting. As a fail-safe learning environment for this purpose, we established an educational brand, *60edu*, for the students to market using the practices they had learned.

The idea of 60edu is simple. As the name already implies to some extent, 60edu is about creating educational videos on various topics that are approximately 60 seconds long. In practice, the hard limit was 90 seconds. The goal of this idea was two-fold: to teach both the video makers and the potential viewers. In order to teach something in mere 60 seconds, the students making the video need to have a solid understanding of what is important about the topic. If it is a topic they have already studied, making such a video about the topic forces them to revise what they have learned, reinforcing the learning experience. This is a concept we plan to utilize in the future as an alternative to e.g. writing short essays.

Thus, after the third lecture, the students were to produce 60edu videos of the topics they had been working on. Once they had produced the videos for the fourth week's lecture, they were to utilize the practices. In order to do so, they were to market the 60edu channel, videos, and the concept using the growth hacking practices they had learned.

For the final week, each student, or group of students, was free to choose a category of growth hacking practices (refer to section 4.1 for the eight categories they chose) to utilize. During the week, they were free to utilize different practices from their category of choice, or at least a minimum of one practice. In this fashion, they were able to get familiar with the tools and platforms required e.g. to carry out display advertising.

At the end of the fifth and final week, each student prepared a presentation for the course end event[3] about their experiences with utilizing the growth hacking practices. In their presentations, the students were asked to (1) teach the audience what the growth hacking practice they used was, and (2) discuss lessons learned actually utilizing the practice.

---

[2] Some of which can be found on Slideshare by searching for "Abrahamsson, P., Himmanen, J."

[3] a recording of which is found on YouTube at https://www.youtube.com/watch?v=EojEsSVNOvk

# 4 TEACHING GROWTH HACKING THROUGH BOARD GAMES

In order to teach growth hacking in a fun and engaging way, we have developed two board games focused on growth hacking and various growth hacking techniques recommended by practitioners (references [1-9, 11, 13-16, 18-19]). Aside from providing an overview of the categories of growth hacking techniques, the board games also offer practical examples of the use of individual growth hacking practices. Both of the games can be downloaded from FigShare, the link to which is in the conclusions section.

## 4.1 Growthopoly

Growthopoly (seen in Fig. 1), as the name indicates, is a Monopoly-inspired board game on growth hacking. In Growthpoly, the players compete against each other with the objective of gaining 5000 followers, and the player to reach that milestone first is the winner.

At the beginning of the game, each player is assigned a player character. Each character specializes in one of eight areas of specialty in growth hacking: (1) Search Engine Optimization, (2) Email Marketing, (3) Social Media Marketing, (4) Public Relations, (5) Product Development, (6) Display Advertising, (7) Content Marketing, and (8) Search Engine Marketing. This character is used as a game marker for moving on the board. In addition to the character, each player chooses a discernible marker for displaying their changing number of followers in the middle of the board.

At the start of each turn, the player whose turn it is rolls a die and advances that many spaces on the game board. The board contains six types of spaces:

- *Growth hacking skill space*. Whenever a player lands on a growth hacking skill space, the player may pay a certain amount of game money to study that skill for a number of turns: one turn for level one, two for level two, and three for level three skills. When the player has learned the skill, they gain the number of followers on the space.
- *Bonus space*. Upon arriving in a bonus space, the player draws a bonus card. Bonus cards are always positive and grant the player with either money, followers, or both.
- *Trade fair space*. In the trade fair, the player may choose to pay a certain sum of money to gain a number of followers instantly.
- *Problem & Solution (prob & solve) space*. The player draws a prob & solve card, which may be either a problem or a solution. Solution cards are used to tackle problems and may be stored for later use, while problems cause immediate, negative effects when drawn unless countered with a solution. Players may trade solutions.
- *The Slush space*. The player spends a maximum of three turns at Slush. At the start of each turn, the player rolls a die to see whether they get more customers or they end up having to leave Slush.
- *The Start space*. Upon arriving in (or simply passing by) the start space after looping around the game board once, the player gains customers and game money.

By learning the different growth hacking techniques for their characters and by landing on the various spaces, players can gain more followers and/or more money. Both are important, but only followers serve as a win condition. If a player lands on a growth hacking skill already learned (owned) by another player, the owner gains the amount of followers listed on the space. If the player lands on the growth hacking skill that is also their player character's specialty, they get twice the amount of followers and learning the skill takes one turn less than it otherwise would.

The game is intended to serve as a general introduction to growth hacking. It teaches the players about the various types of growth hacking practices (e.g. Search Engine Optimization) that are available and can be utilized by startups. It does not contain much educative content as far as micro-level growth hacking techniques go, however. The other game, discussed in the following sub-section, on the other hand focuses specifically on growth hacking techniques or practices by presenting singular practices one at a time.

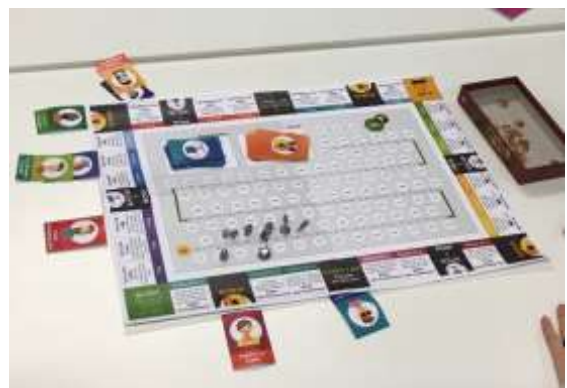

**Fig. 1**. The Growthopoly Game Board

## 4.2 The Game of Growth

In *The Game of Growth*, the players form a team that is intended to emulate a startup organization. Rather than competing against each other as single players, the team then aims to win the game together as a team. The objective of the game is to get 5000 followers for the team's hypothetical software service.

Before beginning, the players choose the type of the startup: tech, service, or entertainment. The team then starts the game with 5000 dollars. Using the 5000 dollars, the players have 10 turns to reach 5000 followers. Each turn is intended to represent one week.

Each turn has three phases, each of which is denoted by drawing a different type of card:

(1) First, at the start of the turn, the team draws an *event* card. The event card applies special rules for that turn (e.g. hiring is cheaper).

(2) Then, the team draws three *hack* cards. The hack cards contain ways to increase the number of followers for the software. For example, a hack card may require the team to pay a few hundred dollars for a chance to gain a few hundred followers by rolling the die favorably. The team may either use or ignore the hack cards, but they are all discarded at the end of the turn either way.

(3) Thirdly, and finally, the team reveals an *employee* card at the end of the turn. The team then either hires or refuses to hire the employee, concluding the turn. Any employees hired by the team will have to be paid a salary at the start of each turn until the end of the game, or until the team fires them. The employees offer various ways for the team to gain more followers.

The game then continues until (1) ten turns have passed, (2) the team has reached the 5000 followers, or (3) the team runs out of money. The game is lost if the ten turns pass without the team reaching the required 5000 followers, or if the team runs out of money before reaching the milestone.

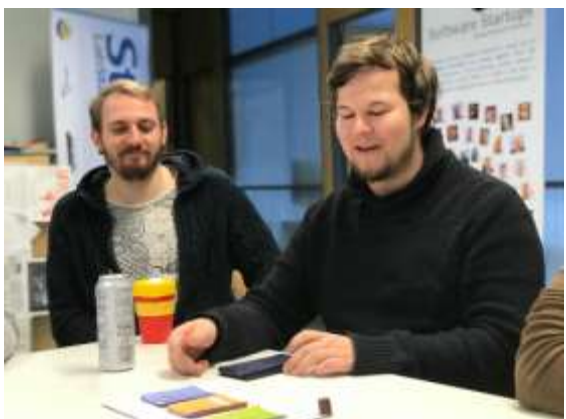

**Fig. 2.** Student Explaining the Game of Growth

The educational value of the game comes primarily from the hack cards. Each hack card contains descriptions of single techniques or practices associated with growth hacking. The cards cover practices such as asking internet celebrities to promote your service, or simply sending personalized emails to targeted prospects ("cold emailing") as a very early-stage startup looking to gain its first users or customers. The game thus takes a more micro-level approach to teaching growth hacking by approaching it from a practical point of view through tangible, real-world practices.

---

[4] https://doi.org/10.6084/m9.figshare.7873847.v2

## 5 DISCUSSION & CONCLUSIONS

In this paper, we have discussed the importance of growth for startups, primarily through the practitioner-oriented construct of growth hacking. To tackle the problem of teaching growth hacking to students, we have presented educative two board games. Both board games can be downloaded from FigShare[4].

The first board game, Growthopoly, offers an overview of growth hacking by categorizing various growth hacking practices into eight categories (e.g. Search Engine Optimization). The second game, the Game of Growth, offers more micro-level insights into growth hacking by focusing on individual growth hacking practices.

Together, these games can serve as a fun way of teaching growth hacking to individuals new to the topic. However, as the games were not empirically validated in this paper, we cannot make claims about their efficiency for teaching growth hacking.

Furthermore, in using games for educational purposes, adverse learning is always a potential issue. Games ultimately have to simplify phenomena due to various restrictions arising from game mechanics. For example, an adverse learning from playing one of the games, The Game of Growth, may be that hiring people is not beneficial because you have to pay them. In the game, employees have a rather small influence on whether or not the game can be won, and thus the game notably downplays the value of additional employees in an organization. A critical approach is needed from the players in order to draw a line between the actually informative content and the game mechanics.

Finally, aside from the board games, we underline the educational value of having had the students utilize the growth hacking practices in a real setting. This became apparent in the lessons learned discussed by the students in the course end event.

For example, the display advertising group learned that:
- If the platform sells display advertising by view count for the advertising, views by bots also eat up the count…
- …thus, limiting the ads to certain geographic locations is important not just to target the right audience, but also to avoid the aforementioned problem with bots
- When advertising e.g. a YouTube channel, it can be beneficial to have the traffic pass through a re-direct link that can provide you with more insightful data about the clicks and clickers of your link

Although the study materials did cover various so-called Dos and Don'ts for the practices, many of the lessons learned were topics not discussed in the literature. Thus, by actually utilizing the practices, the students were able to gain valuable experience not only about the practices but also about the various tools used to carry them out in practice.